\documentclass[]{spie}  

\usepackage{amsmath,amsfonts,amssymb}
\usepackage{graphicx}
\usepackage{siunitx}
\usepackage[colorlinks=true, allcolors=blue]{hyperref}
\usepackage{booktabs}
\usepackage{multirow}
\usepackage{color}
\usepackage{array}
\newcolumntype{P}[1]{>{\centering\arraybackslash}p{#1}}
\newcolumntype{M}[1]{>{\centering\arraybackslash}m{#1}}
\title{Perceptually-inspired super-resolution of compressed videos}

\author[a]{Di Ma, Mariana Afonso, Fan Zhang and David R. Bull}

\affil[a]{Department of Electrical and Electric Engineering, University of Bristol, Merchant Venturers Building, Woodland Road, BS8 1UB, Bristol, United Kingdom}

\authorinfo{Correspondence emails: di.ma@bristol.ac.uk, mariana.afonso@bristol.ac.uk, fan.zhang@bristol.ac.uk, and dave.bull@bristol.ac.uk}

\begin{document} 
\maketitle

\begin{abstract}
Spatial resolution adaptation is a technique which has often been employed in video compression to enhance coding efficiency. This approach encodes a lower resolution version of the input video and reconstructs the original resolution during decoding. Instead of using conventional up-sampling filters, recent work has employed advanced super-resolution methods based on convolutional neural networks (CNNs) to further improve reconstruction quality. These approaches are usually trained to minimise pixel-based losses such as Mean-Squared Error (MSE), despite the fact that this type of loss metric does not correlate well with subjective opinions. In this paper, a perceptually-inspired super-resolution approach (M-SRGAN) is proposed for spatial up-sampling of compressed video using a modified CNN model, which has been trained using a generative adversarial network (GAN) on compressed content with perceptual loss functions. The proposed method was integrated with HEVC HM 16.20, and has been evaluated on the JVET Common Test Conditions (UHD test sequences) using the Random Access configuration. The results show evident perceptual quality improvement over the original HM 16.20, with an average bitrate saving of 35.6\% (Bj{\o}ntegaard Delta measurement) based on a perceptual quality metric, VMAF. 

\end{abstract}


\section{INTRODUCTION}
\label{sec:intro}  

Recently, with the increased requirement for high quality, more immersive video content, the tension between the large amounts of video data consumed everyday and the available bandwidth is ever increasing. Video compression techniques are thus essential tools in resolving this conflict. \cite{bull2014communicating}. 

In the past thirty years, a series of video compression standards have been successfully developed and widely adopted, where each generation introduces new tools and formats. Recent efforts include the ongoing standardisation of  Versatile Video Coding (VVC) \cite{bross2018versatile}, which is expected to achieve 30-50\% compression gains compared to its predecessor, High Efficiency Video Coding (HEVC) \cite{hevc}. Moreover, the Alliance for Open Media (AOM) \cite{aom} has developed  open-source video coding algorithms such as AOMedia Video 1 (AV1) \cite{av1}, which achieves comparable coding performance to standardised codecs such as HEVC. 

Spatial resolution adaptation has begun to play an important role in video compression, initially for scalability in the context of low bit rate applications \cite{shen2011down,georgis2015reduced}. In this type of method, lower resolution video content is compressed by the encoder before transmission, while at the decoder, reconstructed lower resolution video frames are up-sampled to the original resolution for display.  

With the advance of machine learning methods, techniques such as  Convolutional Neural Networks (CNNs), have been employed to enhance  up-sampling performance using intelligent super resolution approaches instead of conventional interpolation filters  \cite{dong2015image,kim2016accurate,ledig2017photo,lim2017enhanced}. Recent results have demonstrated the potential of these methods when integrated into modern codecs \cite{li2018convolutional, lin2018convolutional, afonso2018spatial, afonso2019video}, where spatial resolution adaptation can be applied at Coding Tree Unit (CTU) level or/and at frame level.

In most reported cases, the CNN models employed for spatial resolution up-sampling were trained to optimise pixel-wise difference, despite the fact that this type of metric does not correlate well with perceptual quality on compressed and spatial resolution-adapted content \cite{zhang2018bvi,bampis2018spatiotemporal}. More recently, Generative Adversarial Networks (GANs) \cite{goodfellow2014generative} have been employed for single image super resolution \cite{ledig2017photo,wang2018esrgan}, demonstrating further improvements in subjective quality. These however have not yet been successfully applied in the context of video compression.

Inspired by our previous work \cite{afonso2019video} and GAN-based super-resolution methods \cite{ledig2017photo,wang2018esrgan}, we present a perceptually-inspired spatial resolution approach that employs a modified CNN model (M-SRGAN) to up-sample compressed video frames. This has been trained using a GAN architecture incorporating  perceptual loss functions. Our method has been integrated into a spatial resolution adaptation framework using the HEVC Test Model (HM) 16.20 as the host codec. The results show that the proposed approach achieves significant coding gains against original HEVC HM 16.20 on JVET (Joint Video Exploration Team) Ultra High Definition (UHD) test sequences using the Random Access Configuration, based on perceptual quality assessment.

The remainder of the paper is organised as follows. Section \ref{sec:algorithm} describes the proposed perceptually-inspired spatial resolution adaptation method in detail.  Section \ref{sec:results} presents its compression results and the complexity figures. Finally, conclusions and future work are outlined in Section \ref{sec:conclusion}.

\section{Proposed spatial resolution adaptation method}
\label{sec:algorithm}

\begin{figure}[ht]
\centering
\includegraphics[width=12cm]{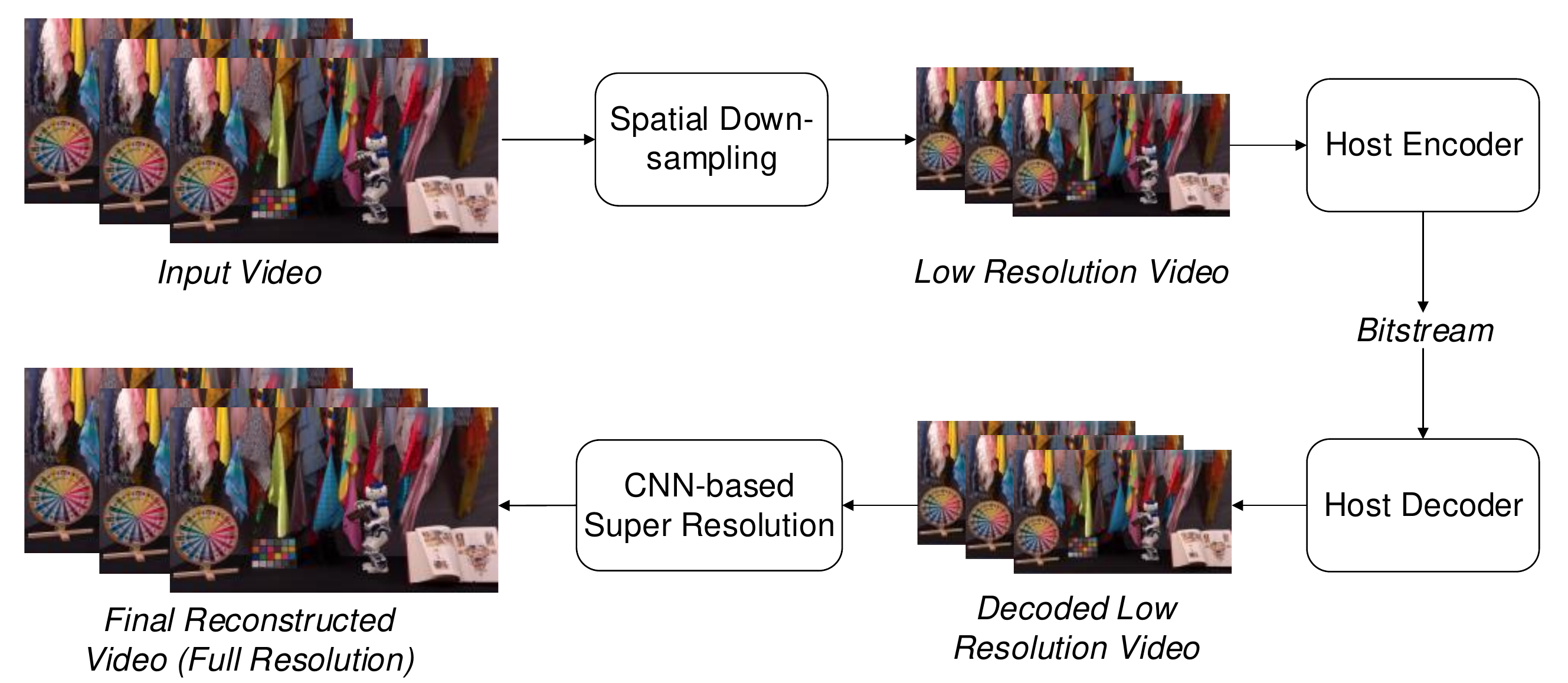}
\caption{Diagram of the proposed perceptually-inspired spatial resolution adaptation method.\label{fig:framework}}
\end{figure}

The proposed adaptation framework is shown in Figure \ref{fig:framework}, in which the input full resolution video frames are spatially down-sampled (by a factor of 2 in this case), and compressed by the host encoder before transmission. At the decoder, the decoded low resolution video frames are up-sampled to their original resolution using a CNN-based super-resolution approach.

\subsection{Spatial Down-sampling}

Original video frames are spatially down-sampled by a factor of 2 using a Lanczos3 filter \cite{burger2009principles}. Compared to other down-sampling filters, such as nearest neighbour and bicubic, it has been reported that Lanczos filters preserve more information in the low resolution version which, in turn, can help reconstructing full resolution video frames at the decoder \cite{afonso2017low}.

\subsection{Encoding and Decoding}

The low resolution video frames are encoded by the host encoder (HM 16.20) prior to transmission. In contrast to our previous work \cite{afonso2019video,afonso2018spatial}, we do not employ resolution adaptation here. Hence the host codec does not need to generate any side information indicating resolution change. This is because, for the vast majority of training sequences at UHD resolution (described in Section \ref{sub:setup}), the proposed method always performs better than the anchor (based on the  perceptual quality metric used) \footnote{This aspect will be further investigated in future work.}. In order to compare with the anchor (original HM 16.20), a fixed QP offset (-6) was applied on base QP values when encoding down-sampled content \cite{afonso2017low,afonso2018spatial,afonso2019video}. This supports the calculation of meaningful Bj{\o}ntegaard Delta results \cite{zhangBD} and the analysis of computational complexity.

\subsection{The CNN Architecture}
\label{sec:CNN}

\begin{figure}[ht]
\centering
\includegraphics[width=17cm]{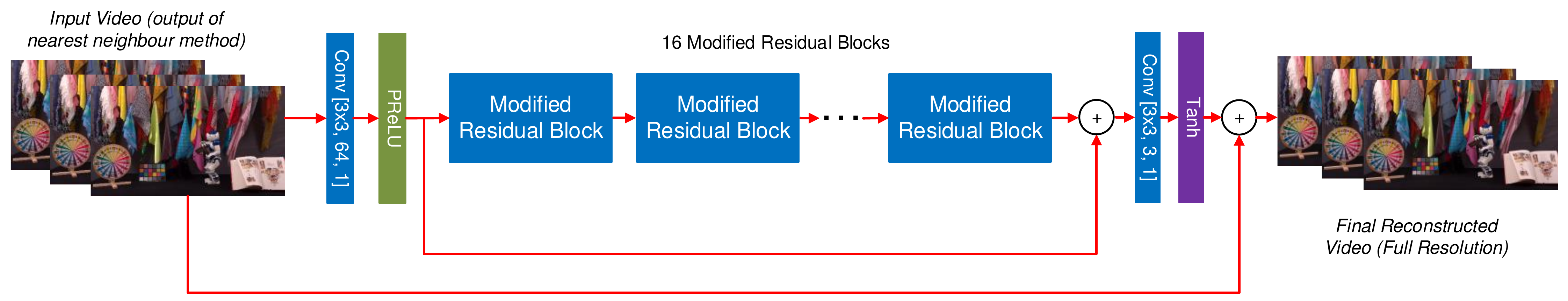}
\caption{Network architecture of the M-SRGAN's Generator (M-SRResNet).\label{fig:generator}}
\end{figure}

\begin{figure}[ht]
\centering
\includegraphics[width=17cm]{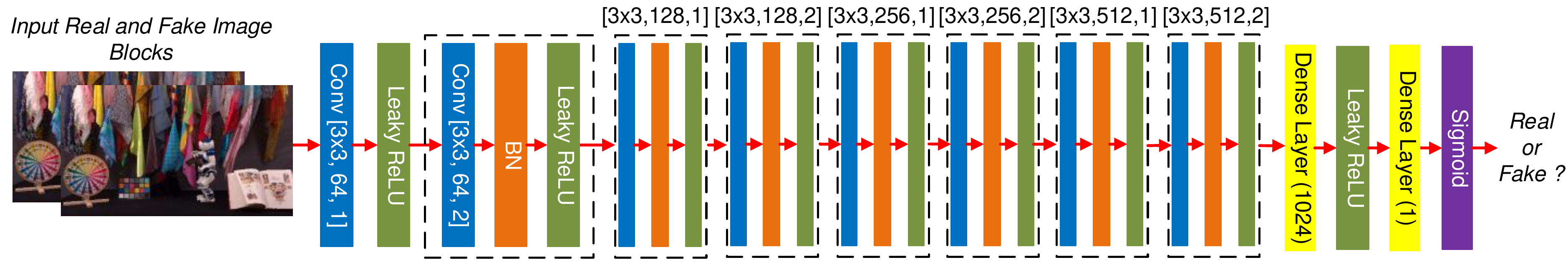}
\caption{Network architecture of the Discriminator for M-SRGAN.\label{fig:discriminator}}
\end{figure}

When the low resolution video frames are decoded, they are first  up-sampled using a nearest neighbour filter to the original resolution\footnote{We have previously shown  \cite{afonso2019video,afonso2018spatial} that nearest neighbour filtering can lead to slightly better reconstruction results.}.  A deep CNN (M-SRGAN) is then employed to further enhance reconstruction quality. This is a  modified version of the SRGAN model \cite{ledig2017photo}. The architectures of the generator (M-SRResNet) and the discriminator are shown in Figure \ref{fig:generator} and \ref{fig:discriminator}.

The generator (M-SRResNet) for the proposed M-SRGAN has a similar architecture to the original SRGAN \cite{ledig2017photo}. The input of the model is 96$\times$96 YCbCr (4:4:4), compressed image blocks (nearest neighbour filter up-sampled), with the target to output colour image blocks close to the original  at the same resolution.

M-SRResNet contains 16 residual blocks after an initial convolutional layer, where each of these has two convolutional layers with a parametric ReLu (PReLu) layer in between. A skip connection is employed in each residual block connecting the input and the output of this block. Two additional skip connections are also used between the output of the initial convolutional layer and that of the 16 residual block, and between the input of the network and the output of the final convolutional layer (the one with a Tanh activation). All the convolutional layers in M-SRResNet employ a kernel size of 3$\times$3 with a stride of 1. The number of channels is 64, except the final convolutional layer which supports 3 feature maps. 

\begin{figure*}[ht]
\centering
\footnotesize
\centering
\begin{minipage}[b]{0.45\linewidth}
\centering
\centerline{\includegraphics[width=0.65\linewidth]{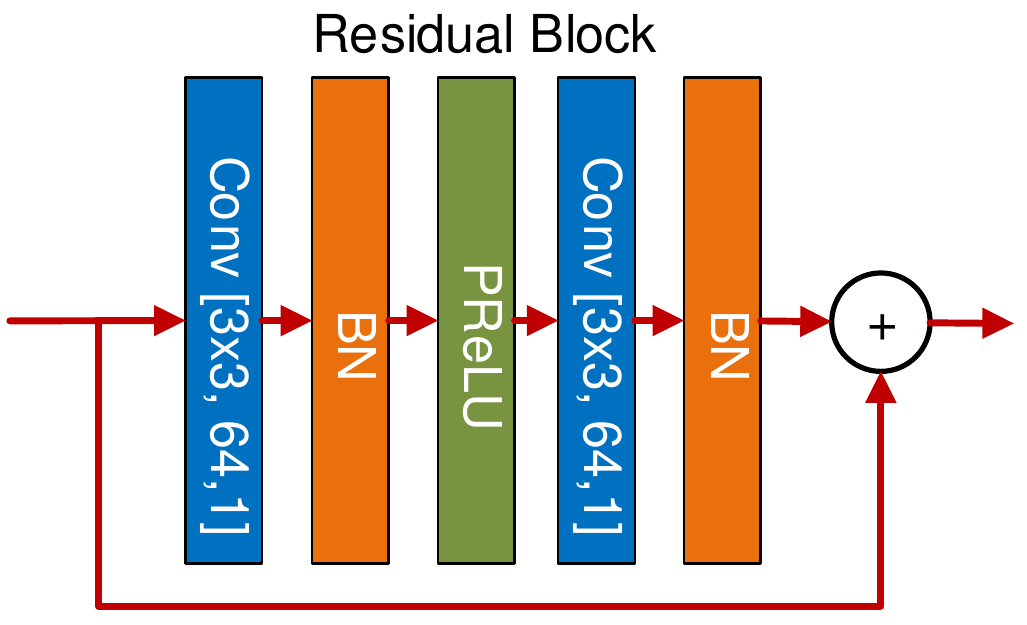}}
\end{minipage}
\begin{minipage}[b]{0.45\linewidth}
\centering
\centerline{\includegraphics[width=0.6\linewidth]{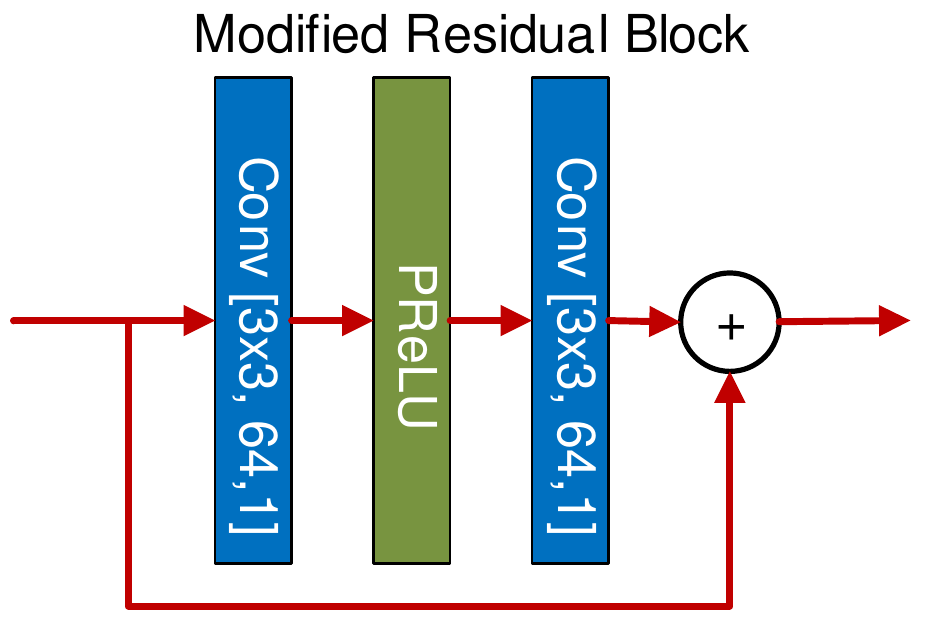}}
\end{minipage}
\caption{(Left) The original residual block in SRGAN \cite{ledig2017photo}. (Right) The modified residual block in M-SRGAN}
\label{fig:block}
\end{figure*}

Comparing to the original SRResNet, the proposed generator does not contain batch normalisation (BN) layers in the residual blocks (as shown in Figure \ref{fig:block}). This is because BN layers were reported to affect the overall network performance \cite{wang2018esrgan} - generating unexpected artefacts - especially for deep GAN networks.

The discriminator in the proposed network has an identical architecture to that of the original SRGAN \cite{ledig2017photo}. Except for the first shallow feature extraction layer, the discriminator concatenates 7 convolutional layers, each of which contains a convolutional operation, a leaky ReLu and a BN. The kernel size of each convolutional layer is 3$\times$3 with a stride of 1 or 2. Different numbers of feature maps are also employed in these layers, from 64 to 512 as shown in Figure \ref{fig:discriminator}. After 7 convolutional layers, 2 dense layers with a leaky ReLu followed by a Sigmoid layer are designed to produce the binary output of the discriminator. 

\subsection{Network Training}
\label{sub:setup}
\subsubsection*{Training Databases.}

 In order to train the proposed M-SRGAN, 108 UHD (3840$\times$2160) source videos were collected from multiple publicly available databases, including BVI-HFR \cite{mackin2015study}, BVI-Texture \cite{papadopoulos2015video}, Netflix Chimera \cite{netflix} and Harmonic Inc \cite{harmonics}. All of these were truncated to 64 frames without any scene cut and converted to 4:2:0 YCbCr format with bit depth of 10. These 108 UHD sequences were then spatially down-sampled to 1080p, 540p and 270p using a Lanczos3 filter to further increase content diversity. This results in a total number of 432 original uncompressed sequences.
 
Each of these sequences was then spatially down-sampled by a factor of 2 and compressed using HEVC HM 16.20 under the JVET Common Test Conditions (CTC) using the Random Access (RA) configuration for four different base QP values: 22, 27, 32, and 37 (a QP offset of -6 was applied). The reconstructed videos, which were up-sampled using the nearest neighbour filter, alongside their corresponding originals were then sub-grouped for four different base QP values, and used as training material for M-SRGAN. This produced four CNN models: 

\begin{equation}
\label{eq4}
{\rm CNN\ Models}=\left\{
\begin{aligned}
{\rm Model_1} & , & {\rm QP\mathrm{\_adjusted}}\leq 18.5\\
{\rm Model_2} & , & {\rm 18.5 < QP\mathrm{\_adjusted}} \leq 23.5\\
{\rm Model_3} & , & {\rm 23.5 < QP\mathrm{\_adjusted}} \leq 28.5\\
{\rm Model_4} & , & {\rm QP\mathrm{\_adjusted}} > 28.5\\
\end{aligned}
\right.
\end{equation}
Here ${\rm QP\mathrm{\_adjusted}}$ represents the adjusted base QP values (after applying the QP offset).

\subsubsection*{Loss Function.}

Loss functions play a crucial role during the training of CNN models. In this work, in order to generate results with improved perceptual quality and realise efficient training, the loss function employed should ideally correlate well with subjective opinions and exhibit relatively low computational complexity. Based on these considerations,  multi-scale structural similarity index (MS-SSIM) \cite{wang2003multiscale} and SSIM \cite{wang2004image} have been employed to train the proposed network\footnote{During the training of the original SRGAN, $\ell$1 (mean absolute difference) and VGG19 \cite{simonyan2014very} were employed in the loss functions. It is however noted that mean absolute difference correlates poorly with subjective results, and a VGG19 based loss function also fails to perform well in the training of M-SRGAN on compressed content.}. MS-SSIM has been previously used to train CNN models similar to M-SRResNet \cite{zhao2016loss}, while SSIM has also been employed in the training of generative adversarial networks to achieve better visual quality \cite{galteri2019deep}.

The training of M-SRGAN consists of two stages. At the first stage, the original MS-SSIM is used as the loss function to train the generator (M-SRResNet). The resulting models are then employed as initial models for the second stage of training. 

At the second stage, in contrast to the original SRGAN \cite{ledig2017photo}, the loss functions from Relativistic GANs (RaGANs) \cite{jolicoeur2018relativistic} have been used to further stabilise the training process and improve the performance of the discriminator network \cite{wang2018esrgan}, which is described below:
\begin{equation}
      \mathcal L_{G}=0.025 \times \ell1+ \mathcal L_{\rm SSIM}+\num{5e-3} \times \mathcal L_G^{Ra}
\label{eq:3}
\end{equation}
\begin{equation}
      \mathcal L_{D}=\mathcal L_D^{Ra}=-E_{x_r}[{\rm ln}({\rm Sig}(C_d(x_r)-E_{x_f}[C_d(x_f)]))]-E_{x_f}[{\rm ln}(1-({\rm Sig}(C_d(x_f)-E_{x_r}[C_d(x_r)])))]
\label{eq:1}
\end{equation}
where $\mathcal L_{G}$ represents the loss function of the generator, while $\mathcal L_{\rm SSIM}$ and $\mathcal L_G^{Ra}$ represent the SSIM loss and adversarial loss of the generator respectively. 
The adversarial loss of the generator $\mathcal L_G^{Ra}$ is defined by (\ref{eq:2}):
\begin{equation}
      \mathcal L_G^{Ra}=-E_{x_r}[{\rm ln}(1-({\rm Sig}(C_d(x_r)-E_{x_f}[C_d(x_f)])))]-E_{x_f}[{\rm ln}({\rm Sig}(C_d(x_f)-E_{x_r}[C_d(x_r)]))]
\label{eq:2}
\end{equation}
where $E$ stands for the mean operation, $x_r$ and $x_f$ are the real and fake image block respectively, and  $C_d(\cdot)$ is the output of the discriminator of M-SRGAN. `Sig' here represents the Sigmoid function. 

In equation (\ref{eq:1}), $\mathcal L_{D}$ and $\mathcal L_D^{Ra}$ are the loss functions for the discriminator and the adversarial loss of the discriminator respectively.

\subsubsection*{Training Configuration}

During CNN training, the input and target frames from each QP group were randomly selected and split into 96$\times$96 blocks, which were also rotated for data augmentation to enhance model generalization. In total, there are more than 100,000 pairs of blocks for each QP group. The CNN model was implemented using the TensorFlow framework (1.10.0) with the following training parameters: Adam optimisation \cite{kingma2014adam} with the hyper-parameters of $\beta_1$=0.9 and $\beta_2$=0.999; batch size of 4$\times$4; 200 training epochs; learning rate (0.0001); weight decay of 0.1 for every 100 epochs.
 
\section{Results and Discussions}
\label{sec:results}

The proposed approach has been integrated with the HEVC reference software (HM 16.20), and was evaluated on UHD sequences from the JVET-CTC dataset \cite{bossen2018jvet} using the Random Access configuration (Main10). The initial base QP values tested were 22, 27, 32 and 37 (before applying a QP offset of -6).

In order to demonstrate the performance of the generative adversarial network and perceptually-inspired loss functions, the performance of M-SRGAN was also compared to that of M-SRResNet (generator only) when the latter was trained using $\ell$1 as a  loss function.

The compression performance of the proposed spatial resolution adaptation framework (using both M-SRGAN and M-SRResNet-$\ell$1 for up-sampling) has been compared with the original HEVC HM 16.20 using the Bj{\o}ntegaard Delta (BD) \cite{BD} measurements based on two quality metrics including PSNR (Y-channel only) and VMAF (Video Multi-Method Assessment Fusion-version 0.6.1) \cite{li2016toward}. PSNR has been widely used for objective quality assessment of compressed video content, although it was reported to correlate poorly with subjective opinions especially on compressed and resolution adapted content \cite{zhang2018bvi,bampis2018spatiotemporal}. VMAF is a machine learning based video quality metric, which was trained on multiple video databases with compressed content at various resolutions \cite{li2016toward,zhang2015perception}.

The computational complexity of the proposed methods was also compared to that of the original HEVC HM 16.20. The encoding was executed on a shared cluster, BlueCrystal Phase 3 \cite{} based in the University of Bristol, which has 223 base blades. Each blade contains 16 2.6GHz SandyBridge cores, with 4 GB RAM. The decoding was executed on a PC which has an Intel(R) Core(TM) i7-4770K CPU @3.5GHz, with 24GB RAM and NVIDIA P6000 GPU device.
  
\subsection{Compression Performance}

\begin{figure}[ht]
\centering
\footnotesize
\centering
\begin{minipage}[b]{0.235\linewidth}
\centering
\centerline{\includegraphics[width=1.05\linewidth]{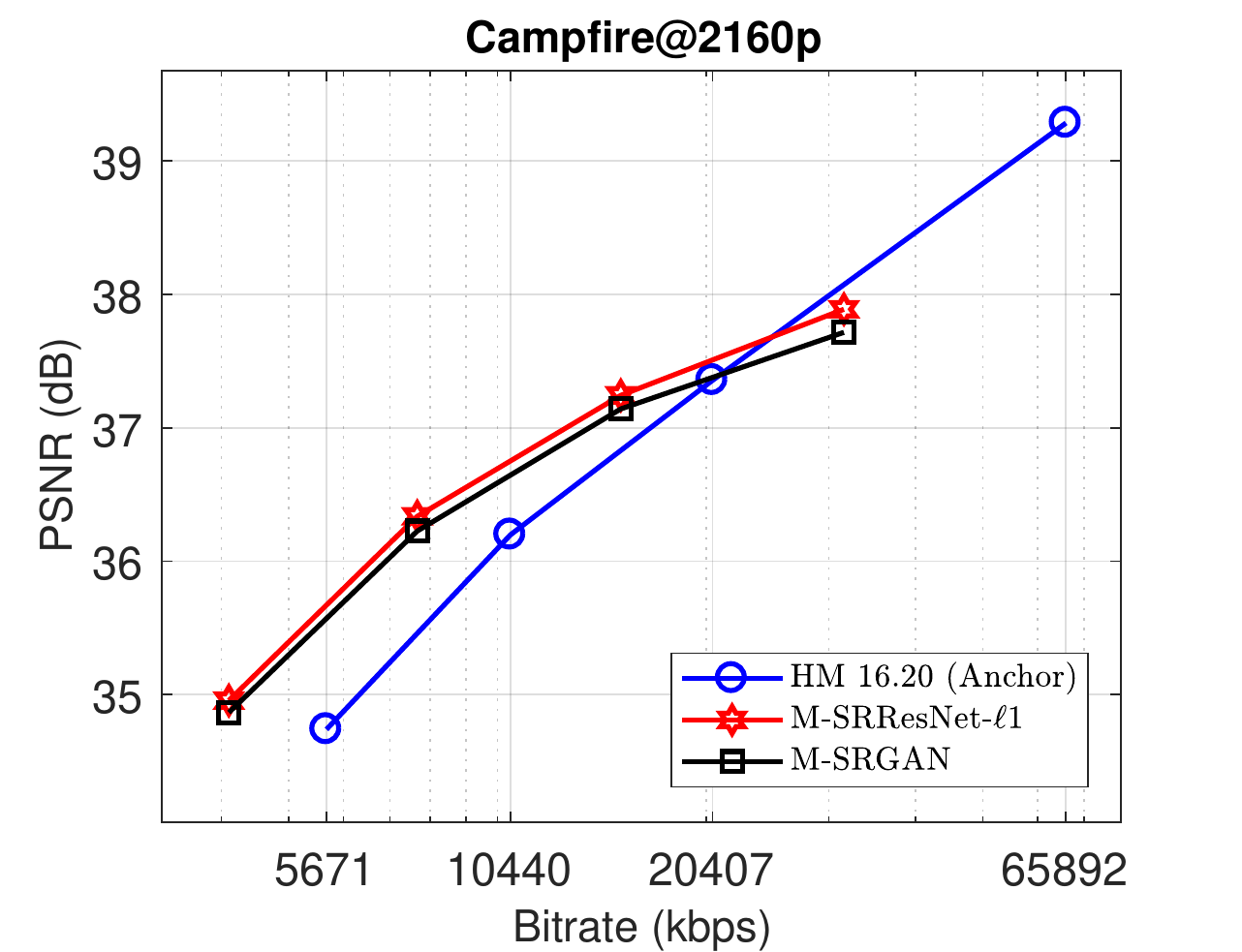}}
\end{minipage}
\begin{minipage}[b]{0.235\linewidth}
\centering
\centerline{\includegraphics[width=1.05\linewidth]{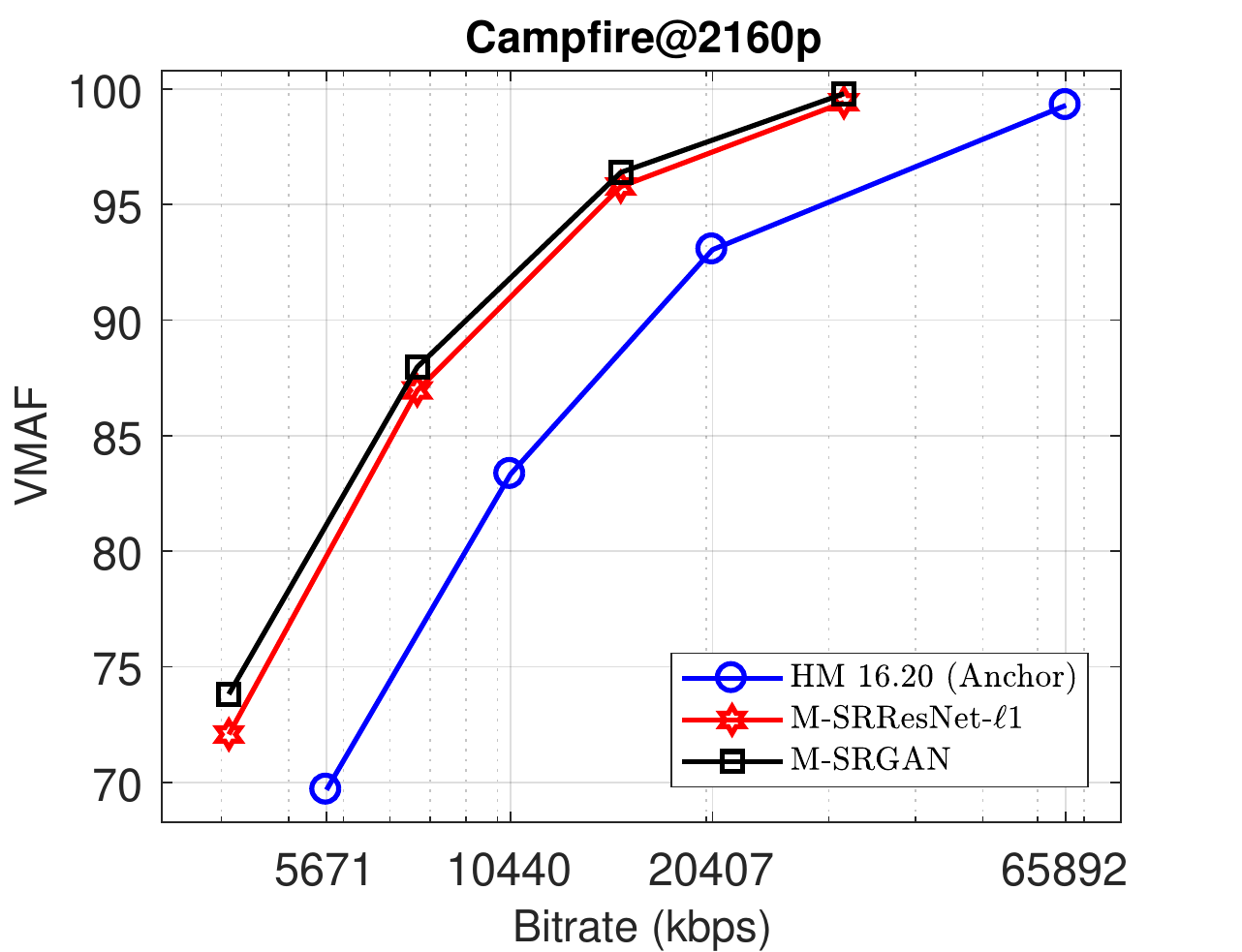}}
\end{minipage}
\begin{minipage}[b]{0.235\linewidth}
\centering
\centerline{\includegraphics[width=1.05\linewidth]{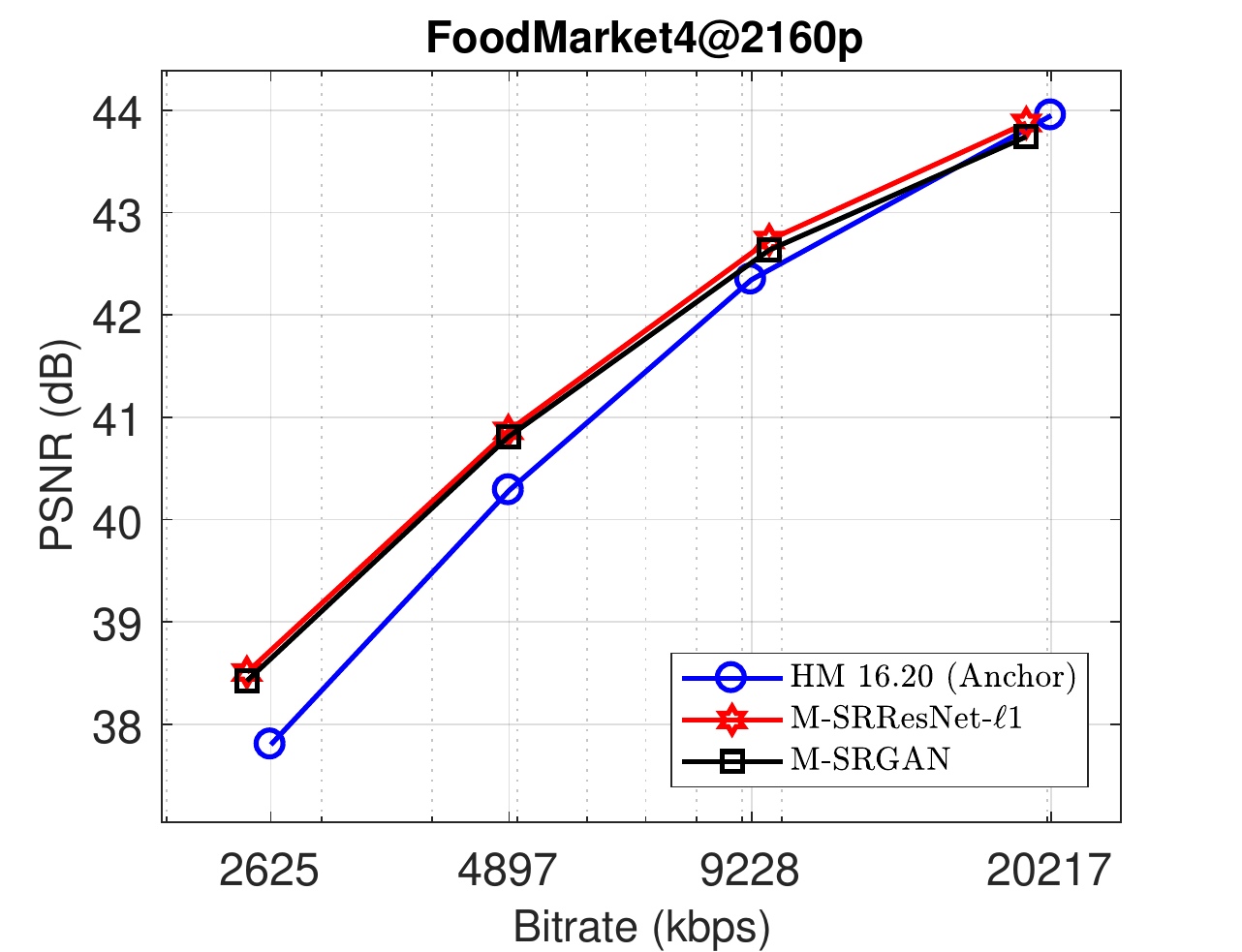}}
\end{minipage}
\begin{minipage}[b]{0.235\linewidth}
\centering
\centerline{\includegraphics[width=1.05\linewidth]{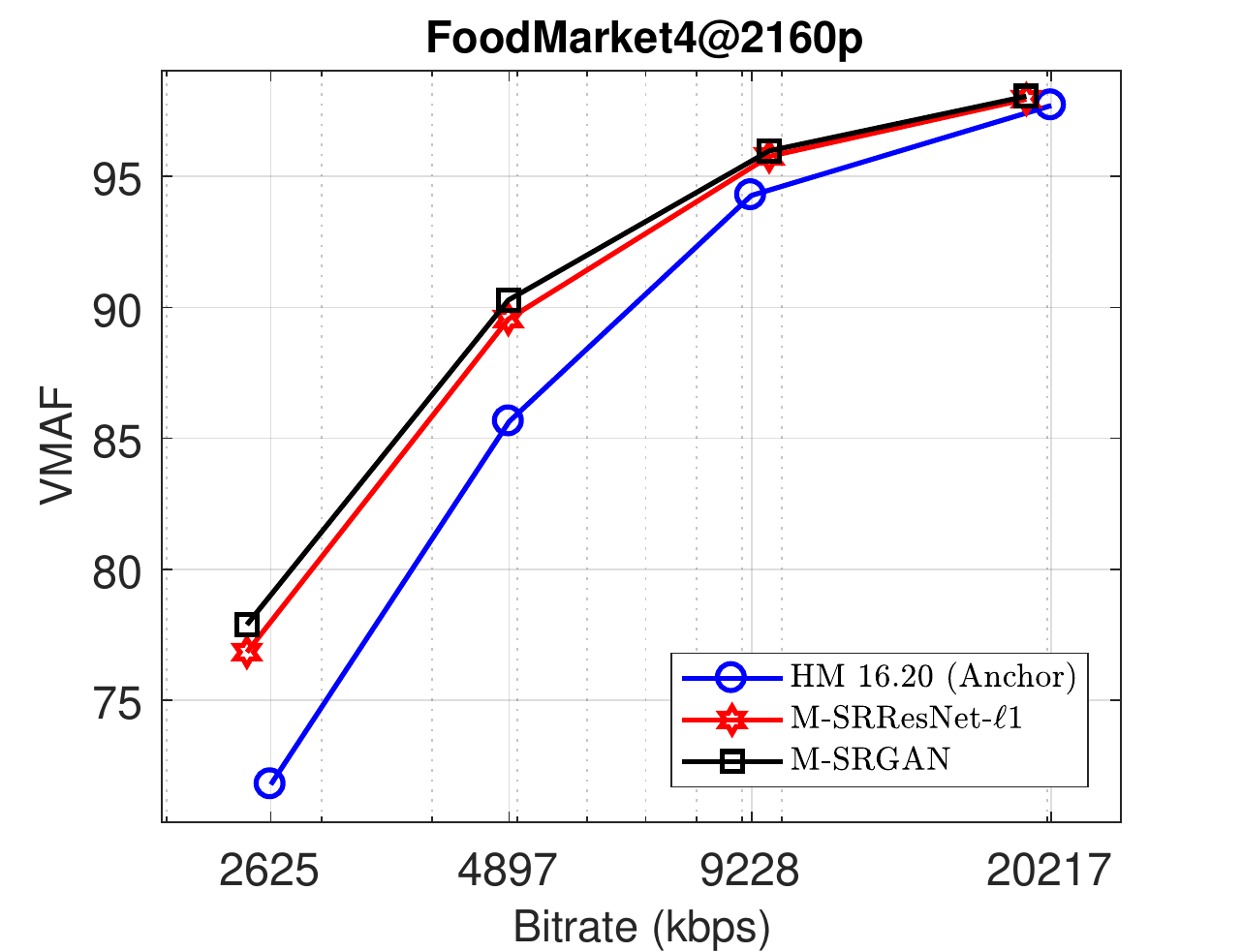}}
\end{minipage}

\begin{minipage}[b]{0.235\linewidth}
\centering
\centerline{\includegraphics[width=1.05\linewidth]{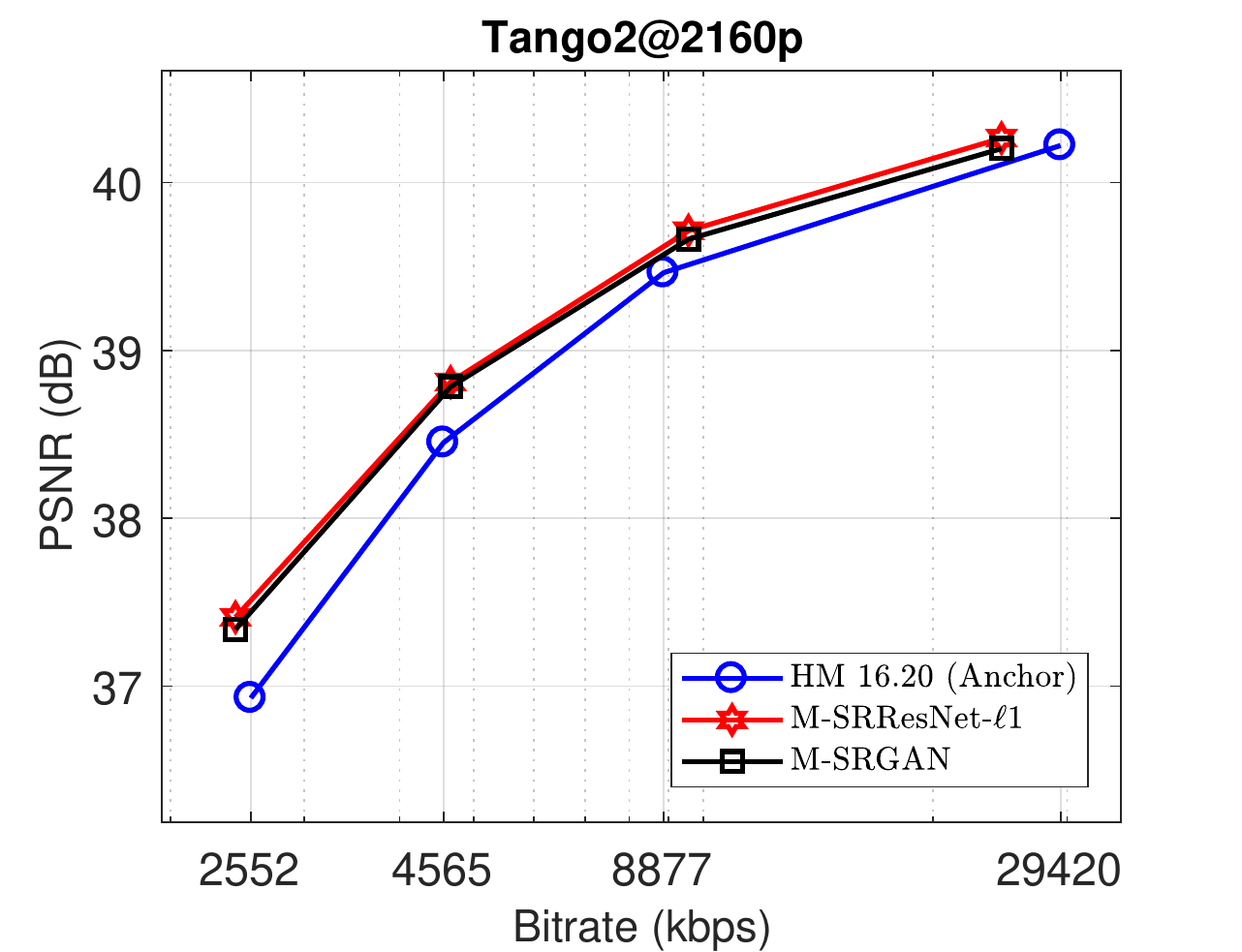}}
\end{minipage}
\begin{minipage}[b]{0.235\linewidth}
\centering
\centerline{\includegraphics[width=1.05\linewidth]{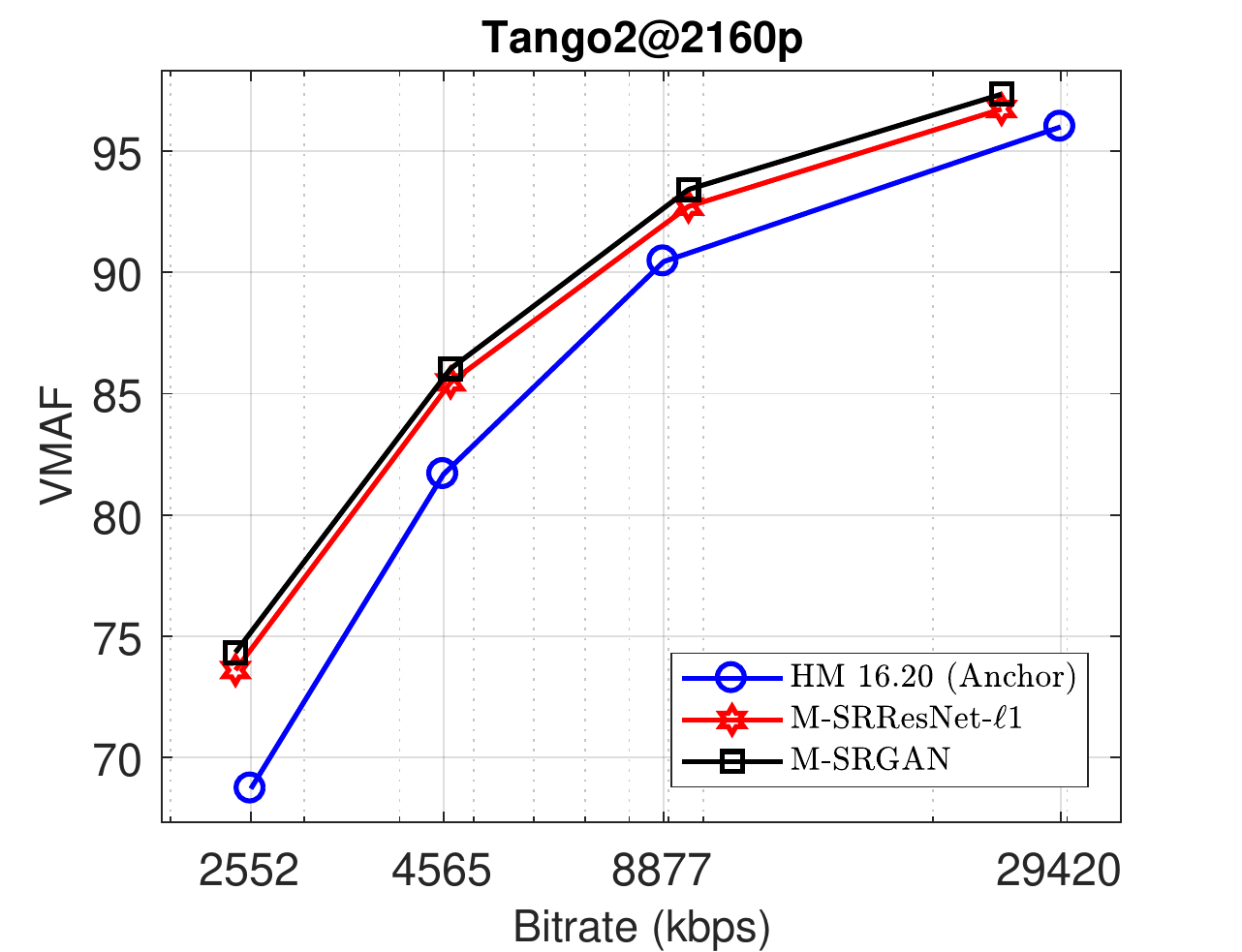}}
\end{minipage}
\begin{minipage}[b]{0.235\linewidth}
\centering
\centerline{\includegraphics[width=1.05\linewidth]{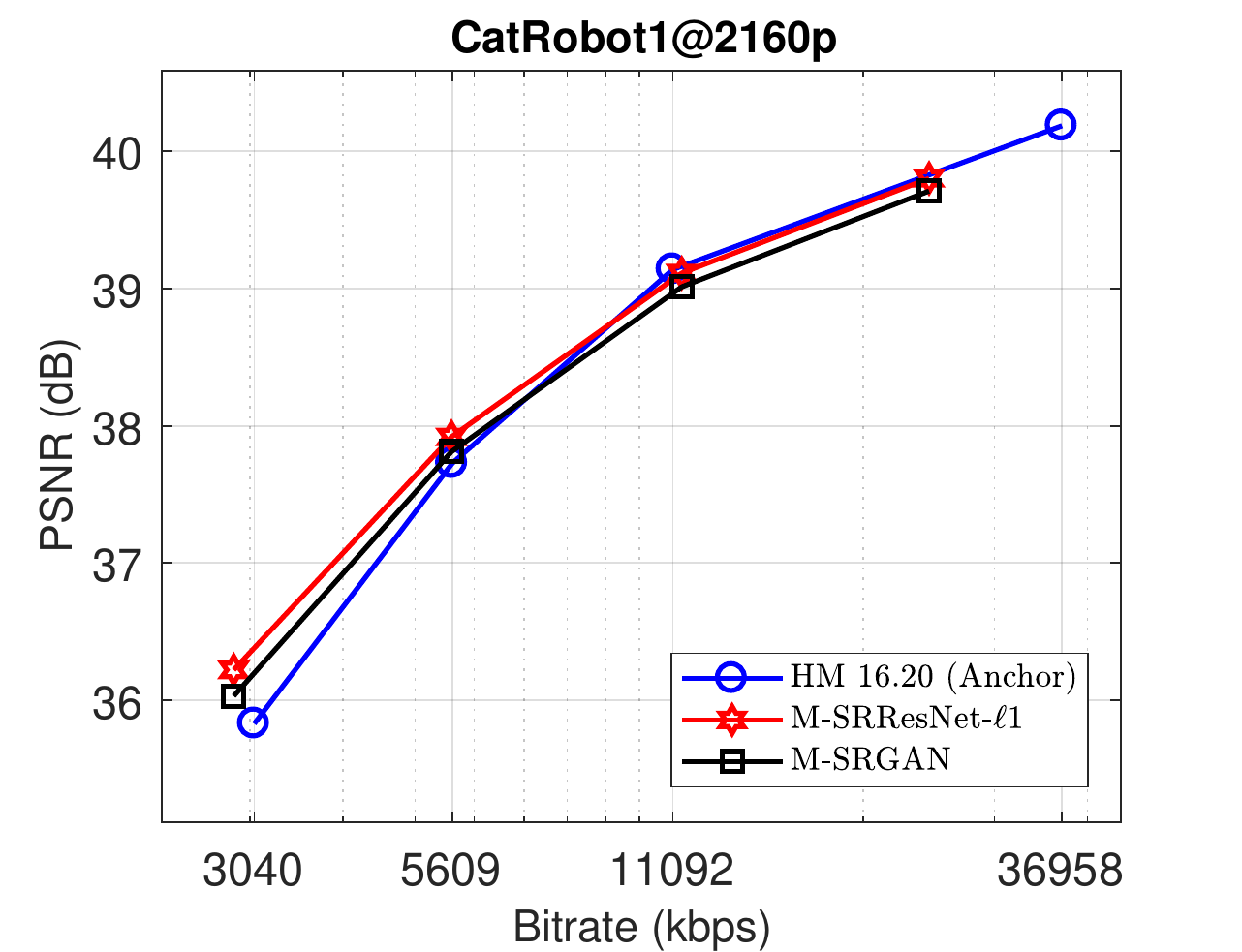}}
\end{minipage}
\begin{minipage}[b]{0.235\linewidth}
\centering
\centerline{\includegraphics[width=1.05\linewidth]{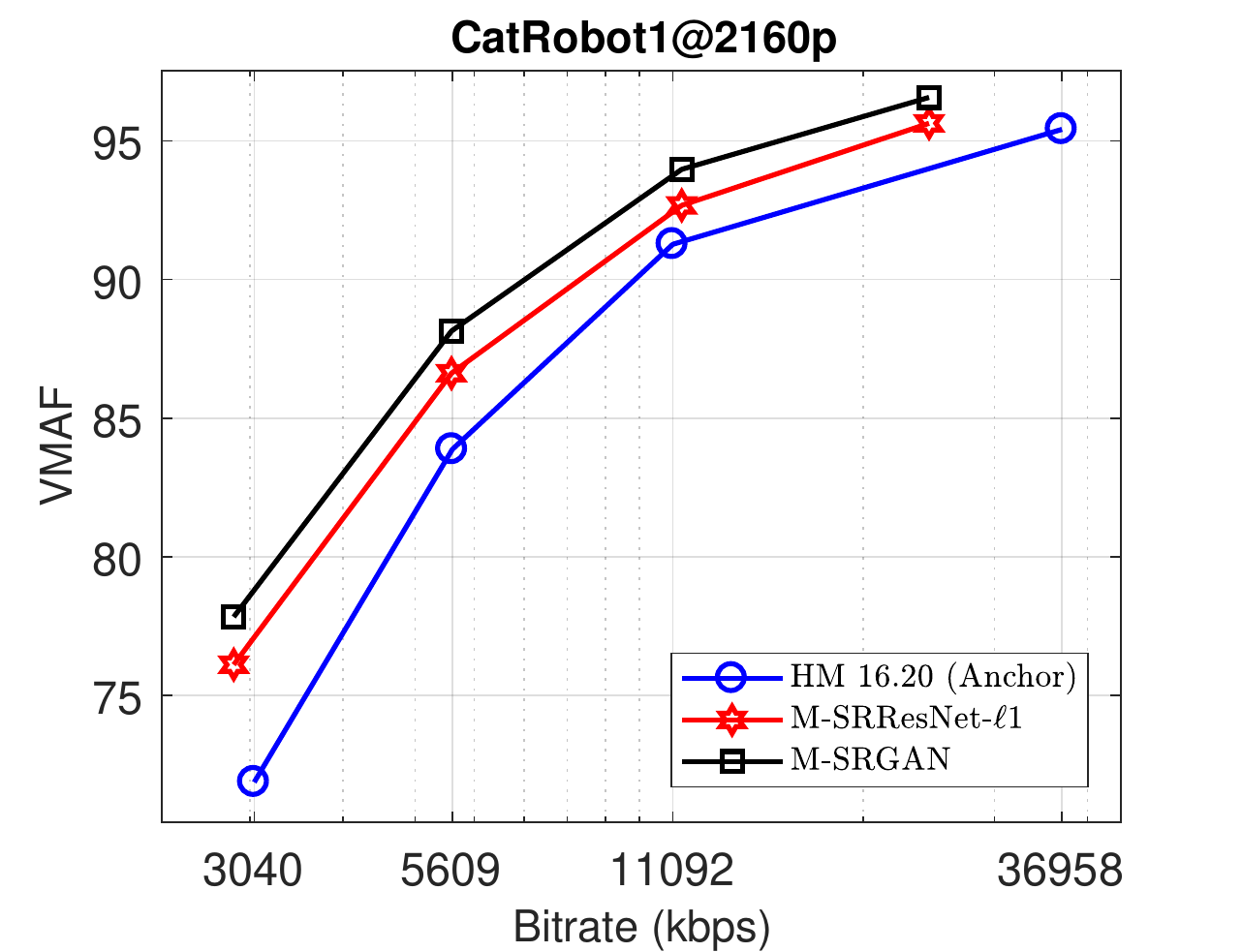}}
\end{minipage}

\begin{minipage}[b]{0.235\linewidth}
\centering
\centerline{\includegraphics[width=1.05\linewidth]{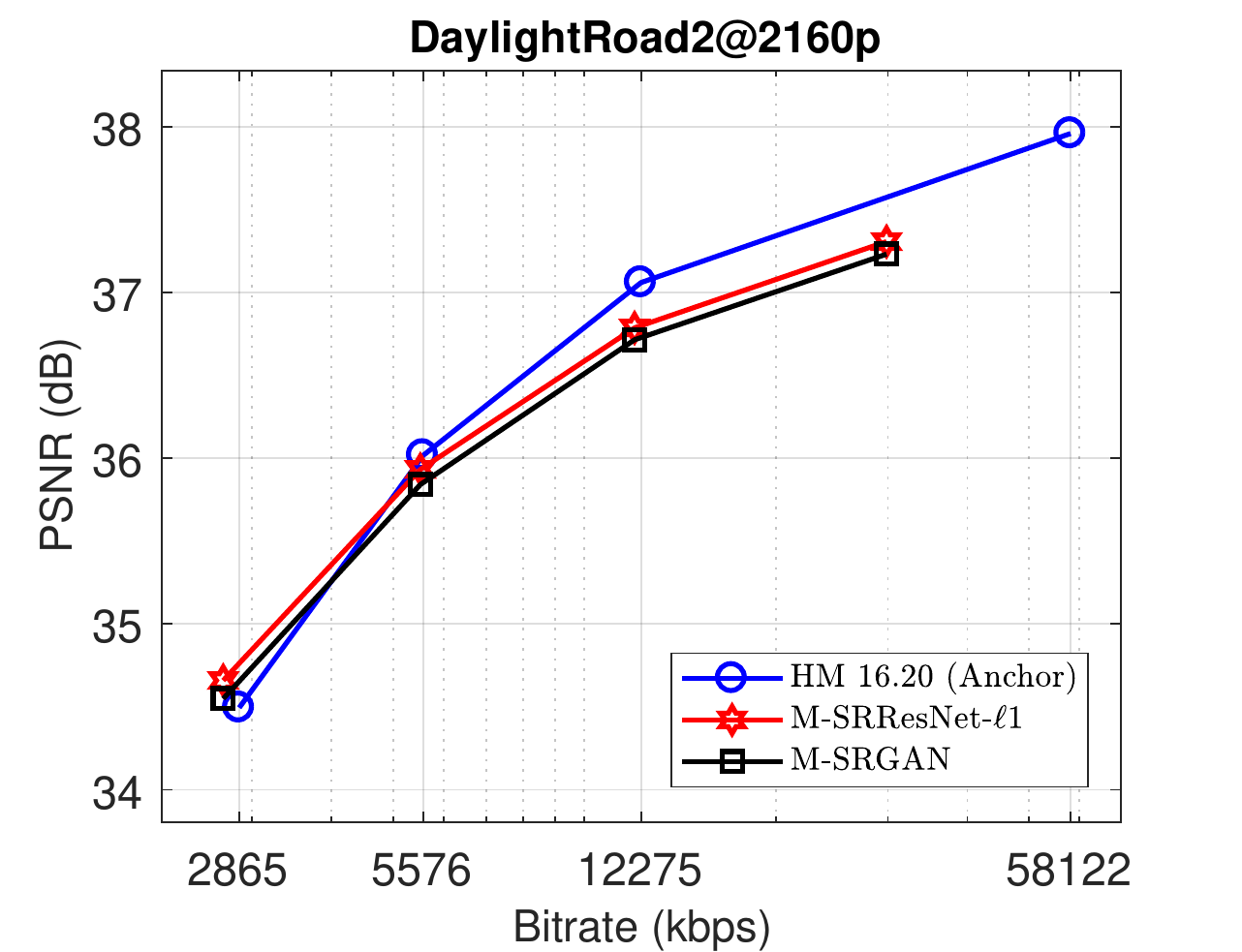}}
\end{minipage}
\begin{minipage}[b]{0.235\linewidth}
\centering
\centerline{\includegraphics[width=1.05\linewidth]{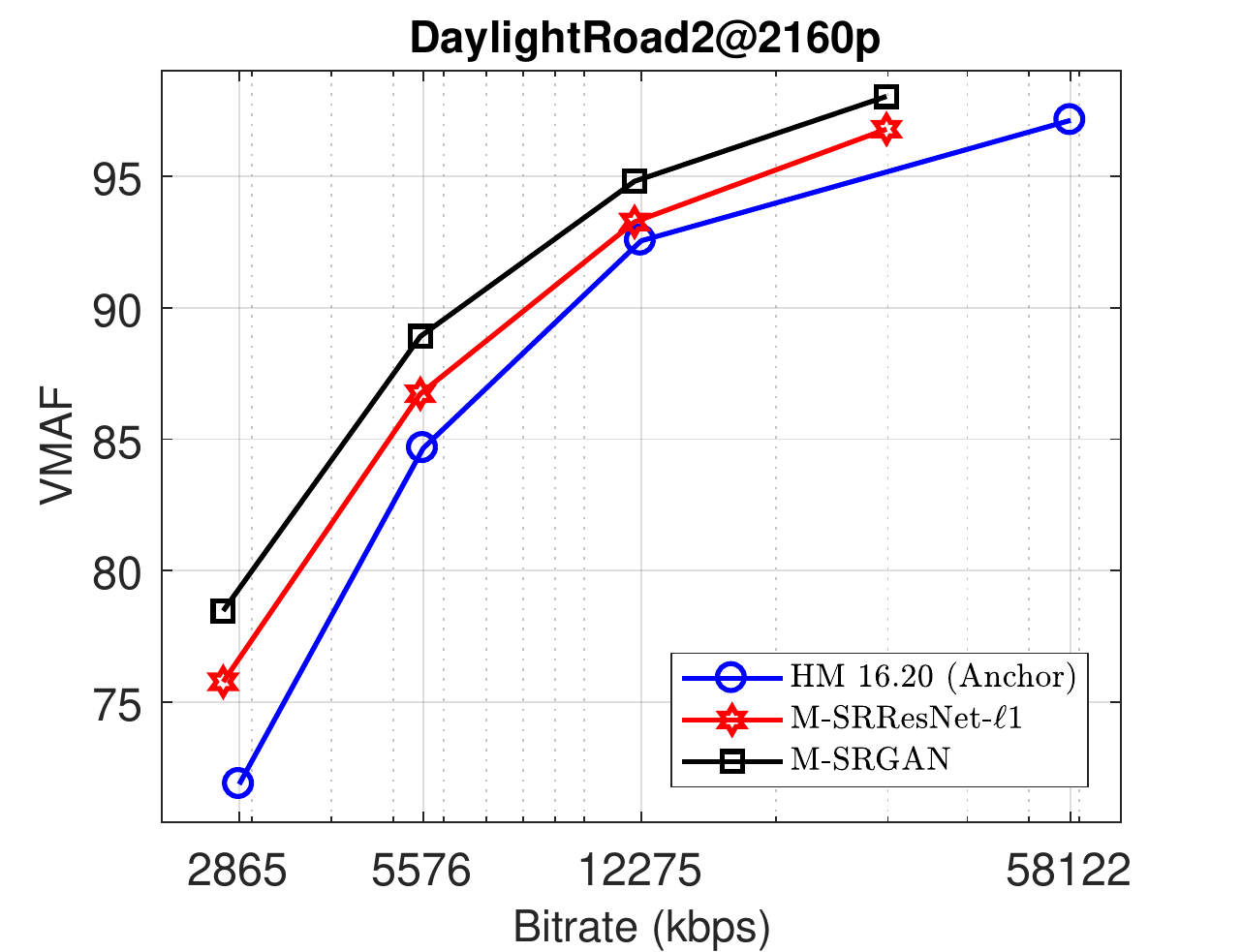}}
\end{minipage}
\begin{minipage}[b]{0.235\linewidth}
\centering
\centerline{\includegraphics[width=1.05\linewidth]{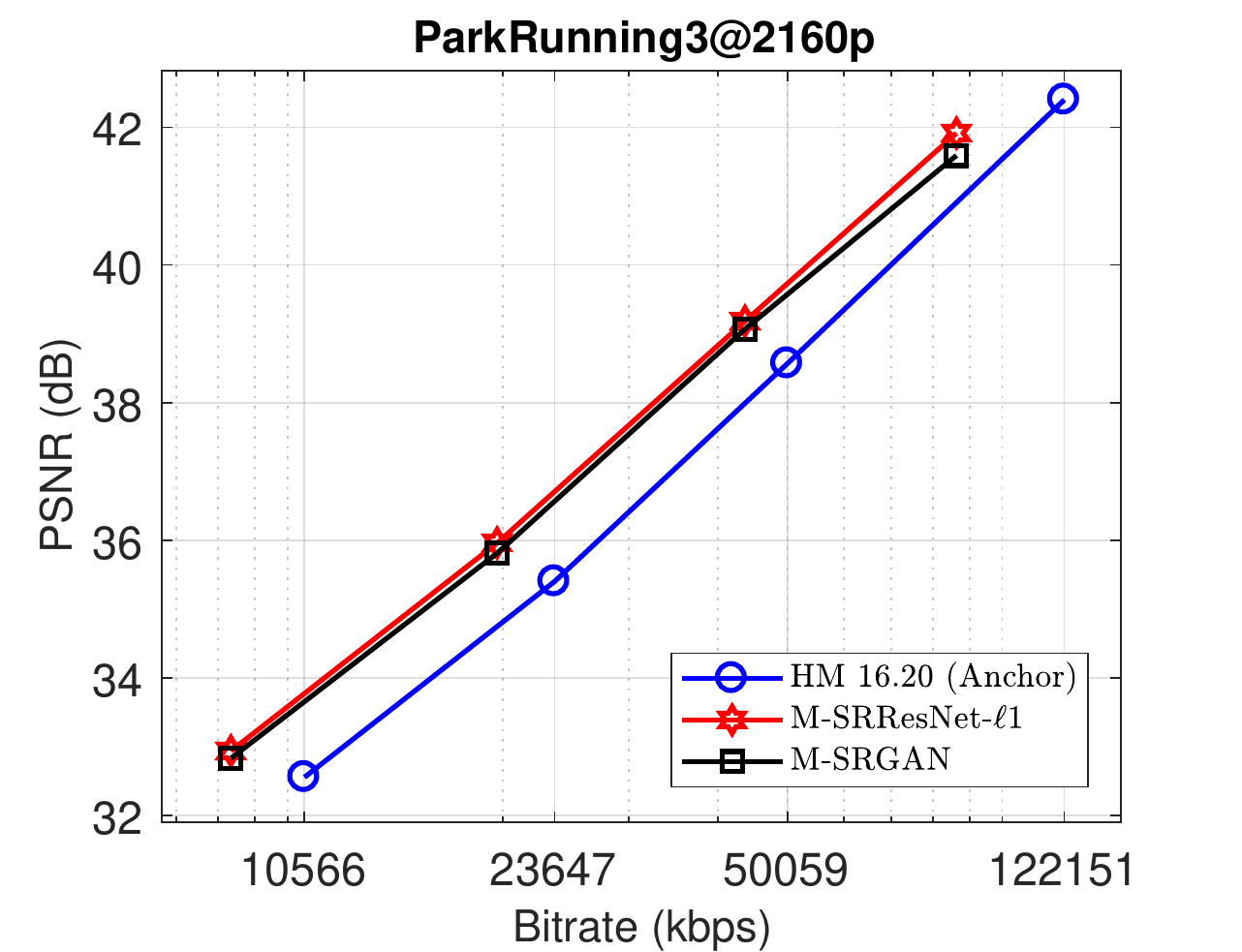}}
\end{minipage}
\begin{minipage}[b]{0.235\linewidth}
\centering
\centerline{\includegraphics[width=1.05\linewidth]{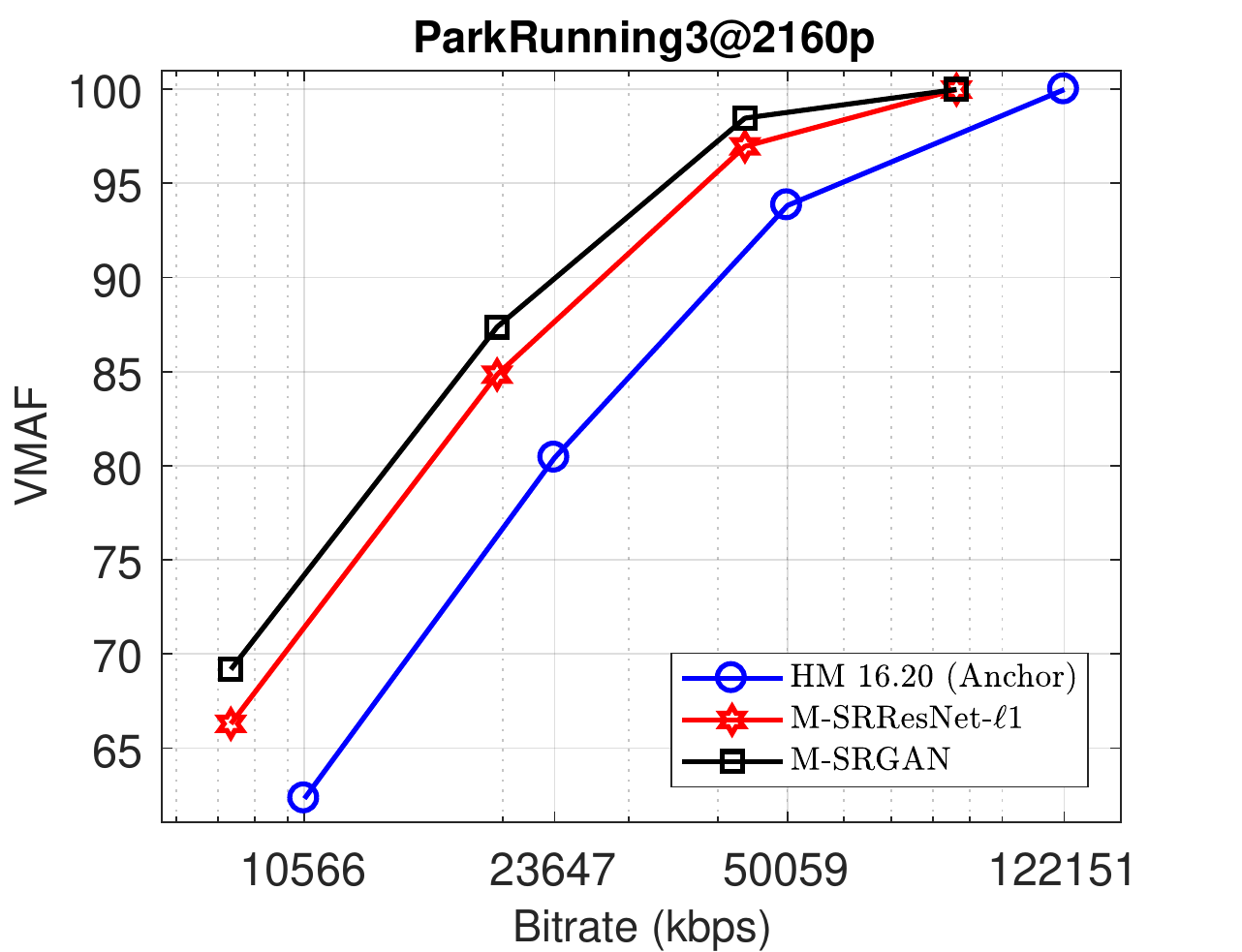}}
\end{minipage}

\caption{Rate-quality curves of the anchor, M-SRResNet and M-SRGAN networks.}
\label{im:curves}
\end{figure}

Table 1 summarises the BD-rate results on the JVET CTC UHD tested sequences, where the proposed spatial resolution adaptation framework (using M-SRResNet-$\ell$1 or M-SRGAN for up-sampling) is  compared to the original HEVC HM 16.20. The rate-quality curves of the anchor HM 16.20, M-SRResNet-$\ell$1 and M-SRGAN are also plotted in Figure \ref{im:curves}. It can be observed that, based on VMAF, both M-SRResNet-$\ell$1 and M-SRGAN offer significant coding gains against the original HM, with average BD-rate gains of -25.9\% and -35.6\% respectively. The improvement is visible for all tested QP cases in Figure \ref{im:curves}. M-SRGAN also achieves an additional 9.7\% savings over M-SRResNet-$\ell$1 due to exploitation of the generative adversarial network and the perceptual loss function.

\begin{figure*}[ht]
\centering
\scriptsize
\centering
\begin{minipage}[b]{0.315\linewidth}
\centering
\centerline{\includegraphics[width=1.01\linewidth]{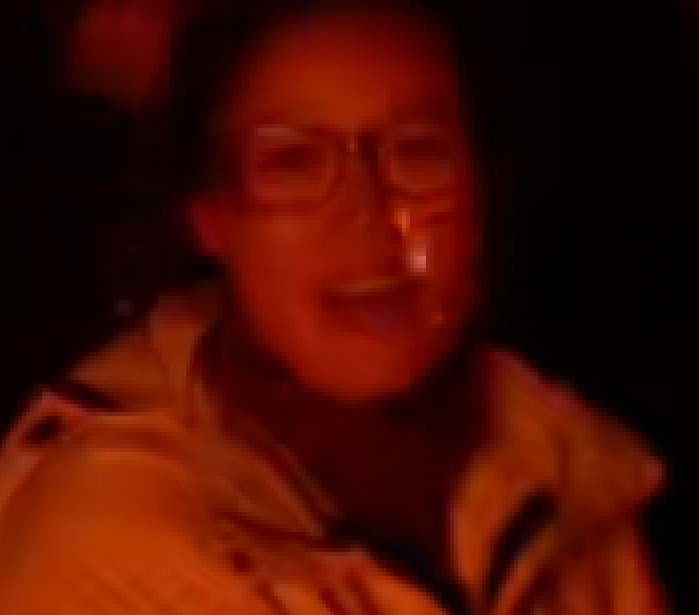}}
(a) Campfire \\ (HM 16.20, QP=37)
\end{minipage}
\begin{minipage}[b]{0.315\linewidth}
\centering
\centerline{\includegraphics[width=1.01\linewidth]{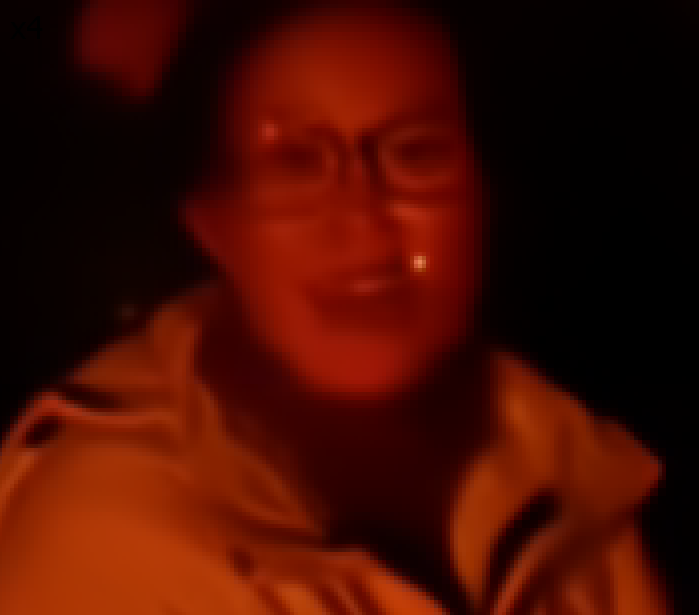}}
(b) Campfire \\ (M-SRResNet-$\ell$1, QP=37)
\end{minipage}
\begin{minipage}[b]{0.315\linewidth}
\centering
\centerline{\includegraphics[width=1.01\linewidth]{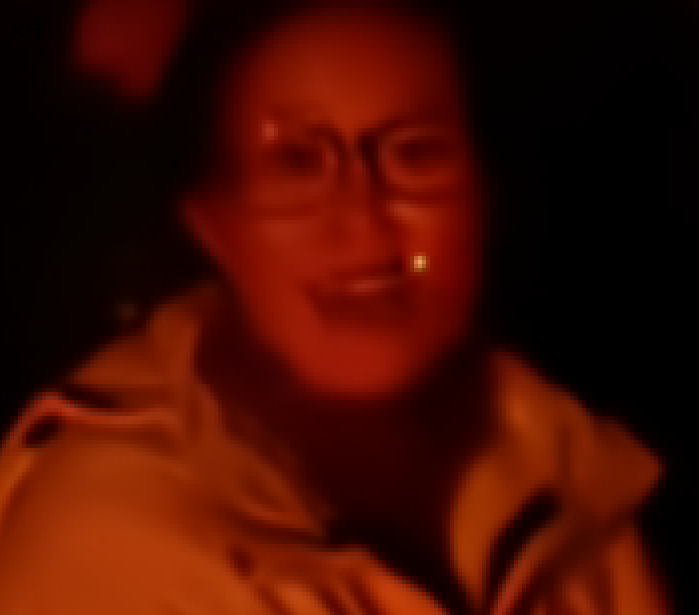}}
(c) Campfire \\ (M-SRGAN, QP=37)
\end{minipage}

\begin{minipage}[b]{0.315\linewidth}
\centering
\centerline{\includegraphics[width=1.01\linewidth]{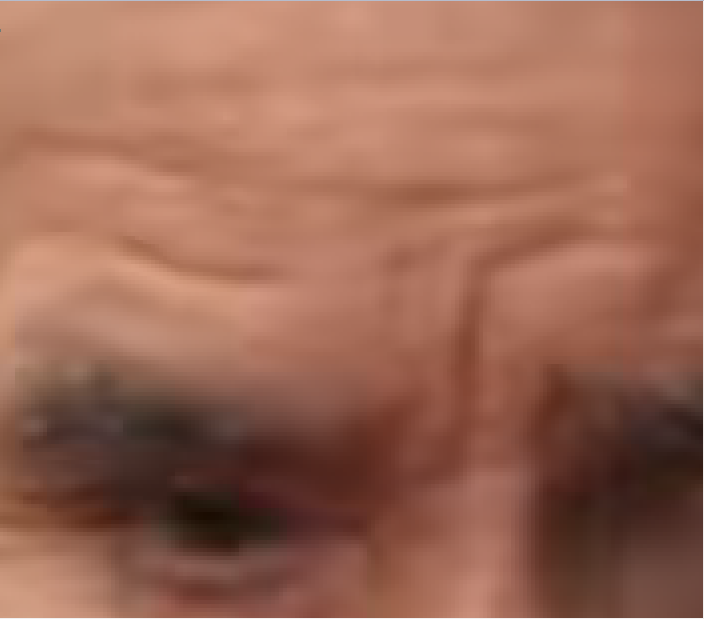}}
(d) Tango2 \\ (HM 16.20, QP=37)
\end{minipage}
\begin{minipage}[b]{0.315\linewidth}
\centering
\centerline{\includegraphics[width=1.013\linewidth]{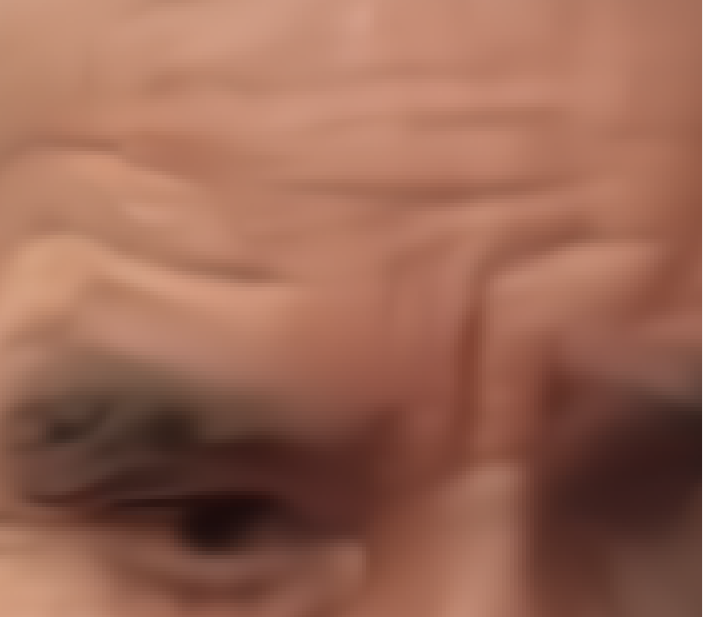}}
(e) Tango2 \\ (M-SRResNet-$\ell$1, QP=37)
\end{minipage}
\begin{minipage}[b]{0.315\linewidth}
\centering
\centerline{\includegraphics[width=1.01\linewidth]{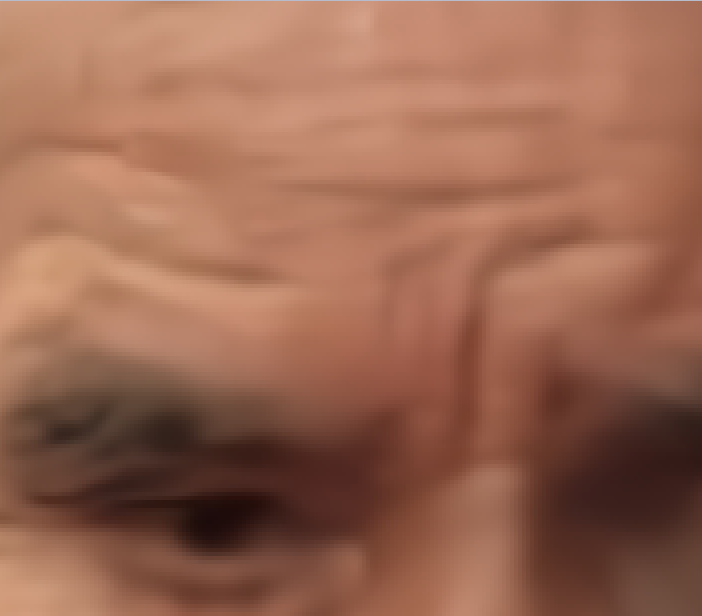}}
(f) Tango2 \\ (M-SRGAN, QP=37)
\end{minipage}

\begin{minipage}[b]{0.315\linewidth}
\centering
\centerline{\includegraphics[width=1.01\linewidth]{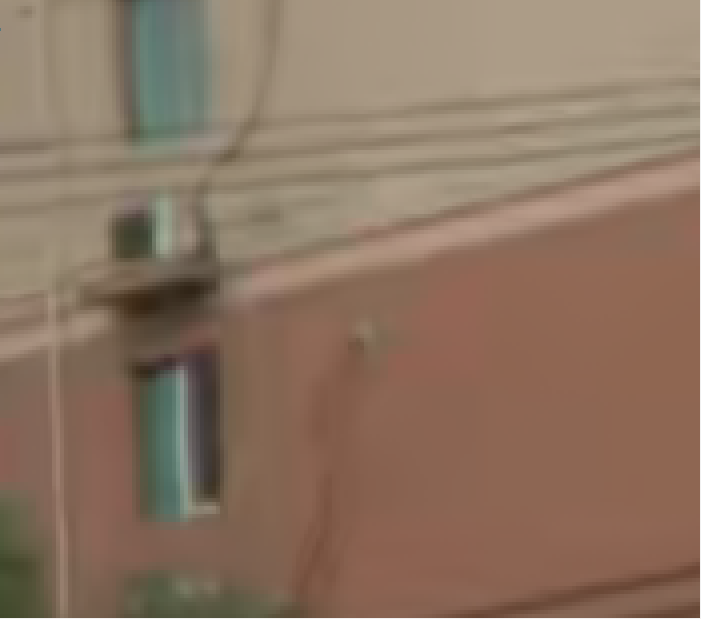}}
(g) DaylightRoad2 \\ (HM 16.20, QP=37)
\end{minipage}
\begin{minipage}[b]{0.315\linewidth}
\centering
\centerline{\includegraphics[width=1.013\linewidth]{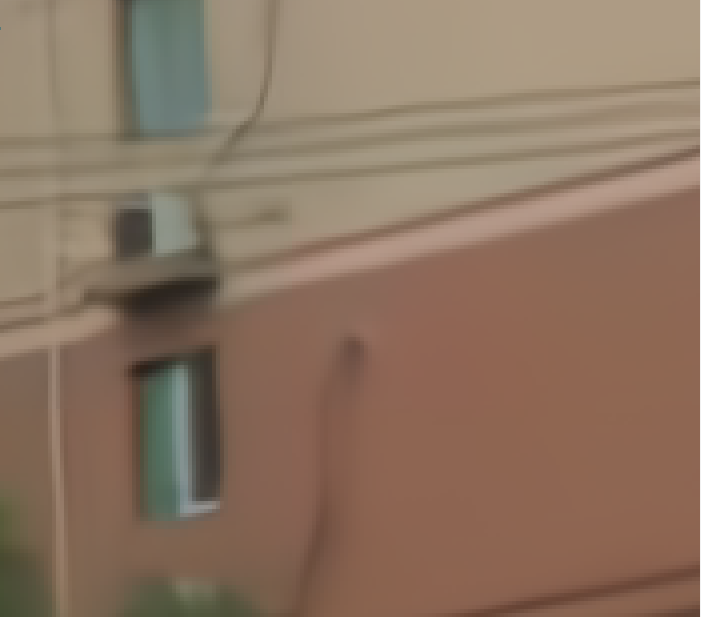}}
(h) DaylightRoad2 \\ (M-SRResNet-$\ell$1, QP=37)
\end{minipage}
\begin{minipage}[b]{0.315\linewidth}
\centering
\centerline{\includegraphics[width=1.01\linewidth]{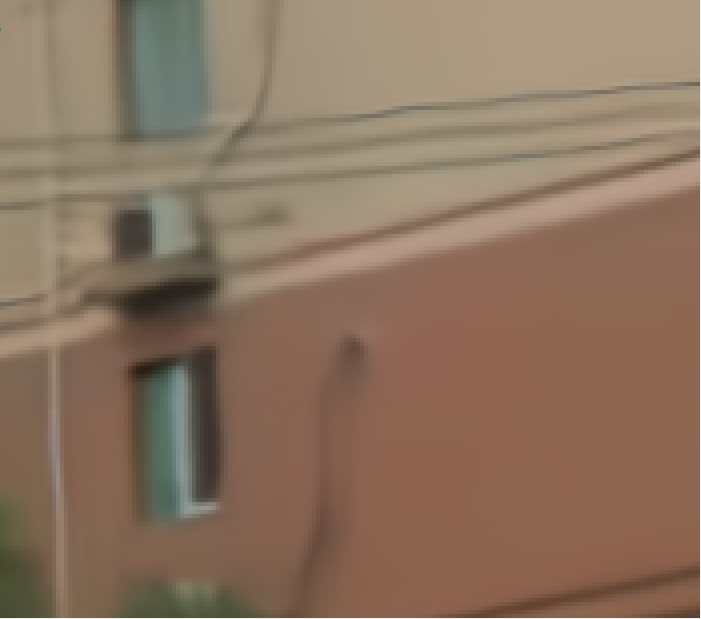}}
(i) DaylightRoad2 \\ (M-SRGAN, QP=37)
\end{minipage}

\caption{Perceptual comparisons between the HM 16.20 and the proposed approach using M-SRResNet-$\ell$1 and M-SRGAN (patches extracted from the 26th, 17th and the 270th frames of `Campfire', `Tango2' and `DaylightRoad2' reconstructed sequences respectively and amplified by 4 times).}
\label{fig:perceptual}
\end{figure*}

The improvement can further be demonstrated by comparing the subjective quality of reconstructed frames. Figure \ref{fig:perceptual} provides a perceptual comparison between the original HM 16.20 and the proposed methods using M-SRResNet-$\ell$1 and M-SRGAN. It is noted that both M-SRResNet-$\ell$1 and M-SRGAN reconstructions exhibit fewer visual artefacts than those of HM 16.20 at similar or even lower bit rates. In addition, M-SRGAN results exhibit slightly more texture detail compared to  M-SRResNet-$\ell$1.

\begin{table}[!t]
\centering
\textbf{Table 1: BD-rate results for JVET CTC UHD tested sequences.}
\begin{tabular}{l |M{3.0cm}|M{3.0cm}|M{3.0cm}|M{3.0cm}}
\toprule
\multirow{2}{*}{Sequence} & \multicolumn{2}{c|}{M-SRResNet-$\ell$1} &
\multicolumn{2}{c}{M-SRGAN}\\
\cmidrule{2-5}
\centering
&   BD-Rate (PSNR) &    BD-Rate (VMAF) &BD-Rate (PSNR)&    BD-Rate (VMAF) \\
\midrule \midrule
Campfire&-26.0\% & -42.0\% &-21.4\% & \textbf{-46.2\%} \\
FoodMarket4&-13.6\% & -22.0\% &-11.2\% & \textbf{-25.9\%} \\
Tango2&-17.0\% & -23.0\% &-13.8\% & \textbf{-27.6\%} \\
CatRobot1&-5.3\% & -22.5\% &-0.2\% & \textbf{-33.5\%}\\
DaylightRoad2&+9.5\% & -20.2\% &+15.5\% & \textbf{-36.4\%} \\
ParkRunning3&-25.9\% & -34.7\% &-23.2\% & \textbf{-44.0\%} \\
\midrule \textbf{Average} & -13.1\% & -25.9\% & -9.1\% & \textbf{-35.6\% }
\\\bottomrule
\end{tabular}
	\end{table}

\subsection{Computational Complexity Analysis}

The average encoding complexity of the proposed approach when integrated into HM 16.20 is only 29.9\% of the original HM 16.20. This is due to the encoding of lower resolution versions of the original content (although with a QP offset of -6). It is however noted that, due to the use of a deep CNN for up-sampling, the average decoding time (the same for M-SRResNet-$\ell$1 and M-SRGAN\footnote{Here only the generator of M-SRGAN is used for evaluation.}) is 27.7 times that of the HM decoder for the tested UHD content.

\section{Conclusions}
\label{sec:conclusion}

In this paper, we have proposed a perceptually-inspired spatial resolution adaptation framework for video compression using GAN-based up-sampling. This approach has been integrated into HEVC HM 16.20 and evaluated on JVET CTC UHD test sequences. The results show significant coding gains for all test sequences based on a perceptual quality metric, VMAF, with an average BD-rate of -35.6\%, accompanied by visible subjective quality improvements. Future work  will focus on the application on lower resolution content, the incorporation of perceptual adaptation strategies and the use of modified network architectures to further improve coding efficiency.
\acknowledgments 
 
The authors acknowledge funding from EPSRC (EP/L016656/1 and EP/M000885/1) and the NVIDIA GPU Seeding Grants.

\bibliographystyle{spiebib} 
\bibliography{refs} 
\end{document}